# Center-to-Limb Variability of Hot Coronal EUV Emissions During Solar Flares

E. M. B. Thiemann[1] (Orcid ID: 0000-0002-5305-9466), P.C. Chamberlin[1] (Orcid ID: 0000-0003-4372-7405), F.G. Eparvier[1] (Orcid ID: 0000-0001-7143-2730), Luke Epp[2]

[1]Laboratory for Atmospheric and Space Physics, University of Colorado, 3665 Discovery Dr., Boulder, CO 80303 (thiemann@lasp.colorado.edu)

[2]Colorado School of Mines, 1500 Illinois St, Golden, CO 80401

## Abstract

It is generally accepted that densities of quiet sun and active region plasma are sufficiently low to justify the optically thin approximation, and it is commonly used in the analysis of line emissions from plasma in the solar corona. However, densities of solar flare loops are substantially higher, compromising the optically thin approximation. This study begins with a radiative transfer model that uses typical solar flare densities and geometries to show that hot coronal emission lines are not generally optically thin. Further, the model demonstrates that the observed line intensity should exhibit center-to-limb variability (CTLV), with flares observed near the limb being dimmer than those occurring near disk-center. The model predictions are validated with an analysis of over 200 flares observed by EVE on SDO that uses 6 lines, with peak formation temperatures between 8.9 and 15.8 MK, to show limb flares are systematically dimmer than disk-center flares. The data are then used to show that the electron column density along the line-of-sight typically increases by $1.76 \times 10^{19}$ cm$^{-2}$ for limb flares over the disk-center flare value. It is shown that CTLV of hot coronal emissions reduces the amount of ionizing radiation propagating into the solar system, and changes the relative intensities of lines and bands commonly used for spectral analysis.

Keywords: Center-Limb Observations; Flares, Spectrum; Spectral Line, Intensity and Diagnostics

## 1. Introduction

Solar flares are widely believed to be the result of magnetic reconnection in the solar corona, which directly heats coronal plasma and accelerates particles along magnetic field lines. These accelerated particles flow towards the field line foot points until local plasma densities are sufficiently thick to stop them, resulting in intense heating and subsequent upwelling of chromospheric plasma towards the loop apices through a process known as chromospheric evaporation. The newly formed loop of dense plasma cools, in part, by radiating its energy into space at extreme ultraviolet (EUV) and X-ray wavelengths. The substantial increase of the loop density that occurs during flares has implications on common assumptions of optical depth typically made when analyzing coronal plasma observations.

Coronal EUV emissions are typically approximated as being optically thin due to the low densities of the non-flaring corona. As such, a coronal source located at disk-center will have the same full-disk integrated irradiance if it were located near the limb instead, assuming all other properties of the disk remain constant. However, since solar flare plasma densities can be one to four orders of magnitude larger than typical quiet sun and active region plasma densities (see Milligan *et al.,* 2012 and references therein), the optically thin assumption will break down, and a flare occurring near the limb may have significantly less EUV irradiance of coronal origin than the same flare occurring near disk-center.

Multi-vantage, spectrally resolved irradiance measurements needed to spectrally characterize solar flare center-to-limb variability (CTLV) do not exist, so statistical analyses of how flares vary spectrally depending on their location on the solar disk must be used instead. This method has been used, for example, to characterize anisotropic emissions of solar flare hard X-rays by McTiernan and Petrosian (1991). Similarly, for EUV emissions, Chamberlin *et al.* (2008) used full-disk integrated solar flare observations from the Solar EUV Experiment (SEE) (Woods *et al.,* 2005) onboard the *Thermosphere Ionosphere Mesosphere Energetics and Dynamics* (TIMED) satellite to show that limb flares have systematically lower EUV spectral content above 35 nm for the same soft X-ray magnitude. Although the 1 nm resolution of TIMED/SEE prevented detailed analysis of the CTLV of individual lines, Chamberlin *et al.* (2008) attributed the differences to the flare spectrum above 35 nm to absorption from chromospheric and transition regions ions, which are assumed to be optically thick due to the relatively high density of these regions.

In this present paper, recent flare spectral measurements made by the *EUV Variability Experiment* (EVE) (Woods *et al.,* 2010) onboard the *Solar Dynamics Observatory* (SDO) are used to characterize the CTLV of hot lines originating in the solar corona. The EVE observations include a data set of over 292 M-Class solar flares measured at wavelengths ranging from 6 to 37 nm with sufficient resolution to characterize the CTLV of individual lines. This study analyzes these data to characterize the CTLV of emissions from six bright flare lines with peak formation temperatures above 7.9 MK. Prior to reporting the observations, Section 2 motivates the data analysis with radiative transfer model results showing that CTLV is expected for hot coronal lines based on the current understanding of flare density and geometry. Sections 3 and 4 present observational evidence of flare CTLV for hot coronal lines. Section 5 discusses the implications of CTLV on flare irradiance, and how spectral analyses of flare plasma assuming the optically thin paradigm are adversely affected.

## 2. Preliminary Analysis

Although non-flaring coronal loops are typically of low enough density that their emissions can be treated as being optically thin, loop densities increase significantly during solar flares, limiting the applicability of the optically thin approximation. To illustrate this, the observed emissions from a series of simple 2-D coronal loops are modeled and analyzed. The modeled loop diameter and foot point separation are 1.8 Mm and 35 Mm, respectively, and consistent with flare loop observations of 41 loops reported by Aschwanden *et al.* (2000). Each point along the loop radiates at unit intensity ($I_0$), and the observed intensity at a specified location ($I_{obs}$) is found by solving the Beer-Lambert radiative transfer equation between each radiating point and the observer,

$$I_{\text{obs}} = I_0 \exp(-\sigma_i \int n_i dz), \ (1)$$

where $\sigma_i$ and $n_i$ are the resonant scattering cross-section and density of the $i^{th}$ ion species, and the integral is taken over the line-of-site along the $z$ coordinate. Note that only resonant scattering is considered and multiple scattering is neglected.

Figure 1 shows model results for the 13.29 Fe XXIII emission line at varying densities and observation angles, where the observation angle is measured from the loop apex. In Panel a, the plasma density is $3\times10^9$ cm$^{-3}$, consistent with observations of active region loops (Aschwanden *et al.,* 2000). In this case, the loop is observed at a 45$^o$ angle; the red mark in the upper right of the figure guides the eye to the observer's location. White indicates that the incident radiance is entirely visible to the observer, while black indicates no emission is visible, and the blue shading bridges these two extremes. The loop in Panel a is entirely visible and the integrated radiance seen by the observer is 97% of the total emitted amount, confirming that the optically thin approximation would be valid for similar active region loops. Panel b is similar to Panel a, but with a plasma density of $8\times10^{10}$ cm$^{-3}$, which is consistent with observations for large solar flares (Milligan *et al.,* 2012); in this case, the loop is viewed from directly above the apex. Again, the entire loop structure is visible, but the underside of the loop is considerably less visible, with the inner-loop regions immediately above the foot-points being completely invisible to the observer. Panel c shows the same loop as Panel b, but here, the observation angle is at 45$^o$. In this case, a substantial portion of the left loop leg is not visible because emissions from it are completely scattered by the right loop leg. These simulations imply that the integrated intensity from a flare loop or sets of loops will depend on an observer's location because portions of the flaring structure may be obscured by other portions of the structure nearer to the observer.

The dependence of observed integrated intensity on observation angle for the loop considered in Panels b and c is shown in Panel d. The vertical axis corresponds with the ratio of the intensity observed directly along the apex axis to the intensity observed from the given angle. It is apparent that the observed integrated intensity above the apex is brighter than that observed off-axis. If this loop were on the Sun with its foot-points on a line parallel with the ecliptic equator, then the angle would correspond with the loop's longitudinal location and observations of this loop would be substantially brighter when viewed at disk-center (0$^o$) than near the limb (*e.g.* 80$^o$). In conclusion, CTLV of hot coronal flare loop line emissions is expected based on typical flare densities and geometries even though the actual configuration is typically much more complex than that considered here. In the following sections, observations of this effect are presented.

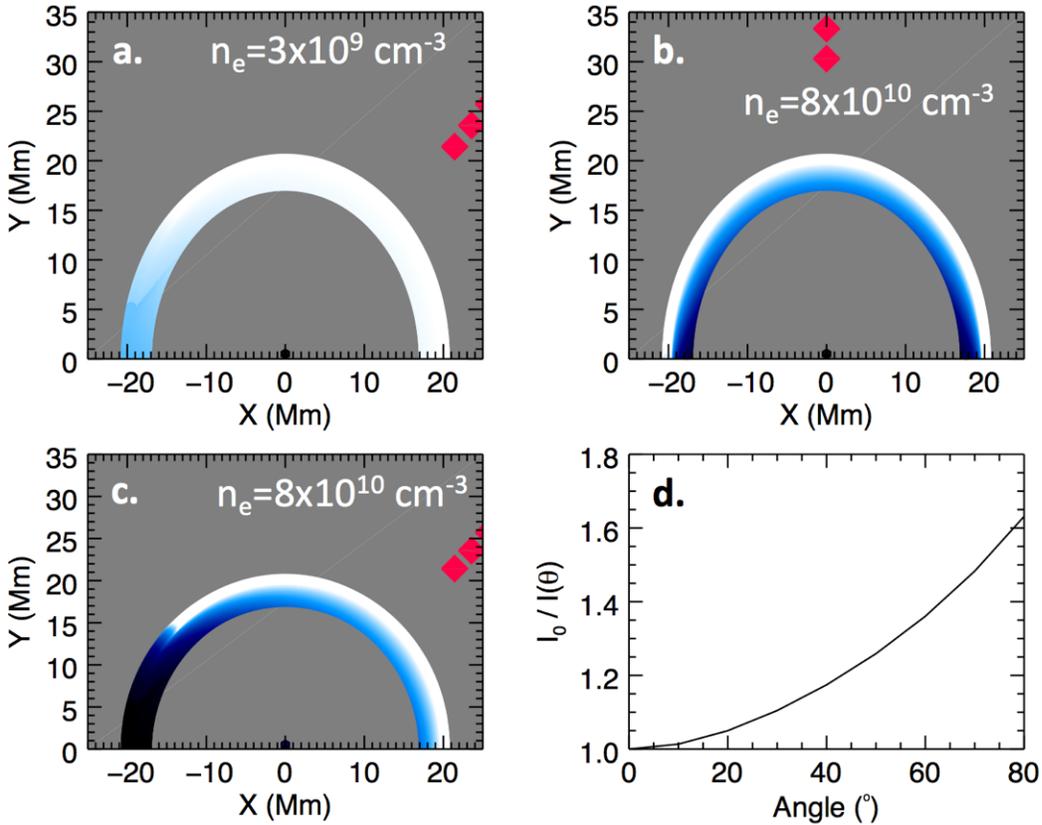

**Figure 1.** Simulated loop radiance as a function of observation angle. a. Modeled radiance from the observation angle indicated by the red mark of a loop of active region density. White coloring indicates the radiance is entirely visible to the observer, while shading indicates the degree to which the radiance is resonantly scattered by plasma along the line of site. b.-c. Same as a. but for flare density and different angles showing that substantial portions of the loop are invisible at the indicated observation points. d. Simulated CTLV as a function of longitude of a flare loop with foot points oriented on a line parallel to the solar ecliptic equator, where CTLV is characterized by the observed integrated intensity normalized by that at disk-center. For example, the loop observed at disk-center is approximately 20% brighter than the same loop observed at 40° longitude.

## 3. Data

### 3.1 Flare Position

Flare positions are identified using SDO *Atmospheric Imaging Assembly* (AIA) (Lemen *et al.,* 2011) images in the 13.3 nm band rendered on the helioviewer.org website, which conveniently displays the angular distance from disk-center to the user-determined cursor position. Flares that appear to have a foot-point over the limb are discarded. The angular distance is converted to solar longitude and latitude. From these coordinates, the orthodromic angle is found between disk-center and the flare location, and this angle is used to classify a flare's location on the disk. Note, the flare location data available in the Solar Events List produced and archived by the National Oceanic and Atmospheric Administration (NOAA) Space Weather Prediction Center (SWPC) are not used because some of the tabulated flare locations are found to be erroneous when independently compared with observations.

## 3.2 EUV Line Irradiance

Solar flare EUV line irradiances are from the SDO/EVE Version 6, Level 2, Spectrum data product, observed by the *Multiple EUV Grating Spectrograph* (MEGS) A channel. MEGS-A made near continuous observations of solar irradiance in the 6 to 37 nm range at 10 second cadence with 0.1 nm spectral resolution and 0.02 nm spectral sampling from 30 April 2010 through 26 May 2014. Over this time, MEGS-A observed thousands of solar flares, including approximately 300 M-class or larger flares.

Solar flare irradiances from the following species (wavelengths) are used in this study: Fe XVIII (9.393 nm), Fe XIX (10.835 nm), Fe XXI (12.875 nm), Fe XXIII (13.291 nm), Fe XXIV (19.203 nm) and Fe XXIV (25.511 nm). These lines are selected because their flare emissions have a high signal to noise ratio, they contribute at least 80% of the observed peak intensity within the line as resolved by MEGS-A, and their peak formation temperature exceeds 9.0 MK, which is within the sensitivity range of temperature ($T$) and Emission Measure (EM) estimates made by the *X-ray Sensor* (XRS) onboard the *Geostationary Operational Environmental Satellites* (GOES) used in the analysis and discussed later in this section.

The flare peak irradiance for each line is identified as follows. Flare line centers and line widths as measured by EVE are identified using the EVE Flare Atlas developed by Hock (2012). Solar flare occurrences are identified using the Solar Events List produced and archived by NOAA SWPC. For each flare, the 13.291 nm Fe XXIII light-curve is plotted, and a time range representing the non-flaring background irradiance level, as well as the time range of the entire flare light-curve are manually identified. An automatic routine then uses these time ranges to identify the flare peaks for the seven lines being analyzed and subtracts the background irradiance. Background subtracted light-curves and selected peak levels are plotted, manually inspected for accuracy, and reprocessed or discarded as necessary if the data appear spurious. Reprocessing or discarding is necessary if, for example, a spurious high data point from an energetic particle triggers a false flare maximum. The background-subtracted peak flare irradiances are used for the analysis discussed in Section 4.

The flare data set is partitioned into 6 subsets according to flare angle. The mean angle (number of flares) for each group are 14.9$^{\circ}$ (45), 31.7$^{\circ}$ (43), 42.7$^{\circ}$ (27), 53.5$^{\circ}$ (31), 61.5$^{\circ}$ (58) and 74$^{\circ}$ (47). The subset of flares with a mean angle near 14.5$^{\circ}$ is treated as the disk-center flare subset against which the other 5 subsets are compared.

Of the initial 292 flares considered, 64 flares are discarded after automated processing. Reasons for discarding are unphysically high (for example, due to unfiltered particle hits) or negative values (for example, due to spuriously high subtracted background values), or missing or unphysical corresponding GOES emission measures. The 228 flares analyzed in this study are listed in Table A1 of the Appendix, which gives the date (in year-day format), hour, soft X-ray magnitude and center-to-limb angle of each flare.

## 3.3 GOES Soft X-rays

Soft X-ray measurements from GOES XRS (Bornmann *et al.,* 1996) are used to determine the flare emission measure. XRS measures irradiance in the 0.1-0.8 nm (Long) and 0.05-0.4 (Short) bands. Prior to analysis, flare plasma emissions are isolated by first subtracting the non-flaring background. These times are taken to be the same as those determined in Section 3.2 for the Fe XXIII emission. The method of Thomas, Starr and Crannell (1985) is then used to estimate flare temperature from the ratio of the two bands. The flare temperature is used with assumed contribution functions and the Long-band irradiance to estimate the emission measure. Note, the assumed spectra and contribution functions used by Thomas, Starr and Crannell (1985) have been updated by White *et al.* (2005), and are incorporated into SolarSoft (Freeland and Handy 1998) as part of the "goes" IDL class, which is used in this study to retrieve XRS band irradiances and derived temperatures and emission measures.

### 3.4 Atomic Physics and Plasma Parameters

Atomic physics and plasma parameters are retrieved from the CHIANTI database, version 8.5 (Del Zanna *et al.,* 2015). Table 1 shows the wavelength, species, peak formation-temperature ($T_{X,\mathrm{pk}}$), degeneracy ratio- Einstein A coefficient product ($g_i/g_j$ x $A_{ij}$), elemental abundance fraction ($A_i/A_H$), ionization fraction ($F_I$) and resonant scattering cross-section for the 6 lines used in the analysis described in Section 4. The 13.3 nm Fe XXIII and 9.4 nm Fe XVIII lines are both blended with Fe XX, and $g_i/g_j$ x $A_{ij}$ for these lines is an average of both species, weighted according to the relative line intensity contributions.

Table 1. Atomic physics and plasma parameters. See Section 3.4 for details.

| Wavelength (nm) | Species | $T_{X,\mathrm{pk}}$ (MK) | $g_i/g_j$ x $A_{ij}$ (GHz) | Abundance (ppm) | $F_I$ | Cross-section (x$10^{-16}$ cm$^2$) |
|---|---|---|---|---|---|---|
| 9.393 | Fe XVIII | 8.9 | 34.69 | 39.8 | 0.35 | 9.19 |
| 10.835 | Fe XIX | 10.0 | 35.84 | 39.8 | 0.26 | 11.8 |
| 12.875 | Fe XXI | 12.6 | 11.67 | 39.8 | 0.22 | 5.74 |
| 13.291 | Fe XXIII | 14.1 | 23.26 | 39.8 | 0.2 | 9.97 |
| 19.203 | Fe XXIV | 15.8 | 8.70 | 39.8 | 0.27 | 12.6 |
| 25.511 | Fe XXIV | 15.8 | 3.44 | 39.8 | 0.27 | 11.7 |

## 4. Methods

The optical intensity of the line emission corresponding with the *i-j* transition ($I_{ij}$) observed through an absorbing medium is related to the intensity at the source ($I_{0,ij}$) by the Beer-Lambert radiative transfer equation,

$$I_{ij} = I_{0,ij}\mathrm{e}^{-\tau}, \qquad (2)$$

where $\tau$ is the optical depth of the medium.

The flare CTLV is characterized by quantifying the difference in $\tau$ for flares observed at disk-center versus those observed near the limb as follows: The flare peak emission measure (EM$_{\mathrm{pk}}$) is related to the source line intensity and peak contribution function ($G_{ij}(T_{\mathrm{pk}})$) according to,

$$I_{0,ij} = G_{ij}(T_{\mathrm{pk}})\mathrm{EM}_{\mathrm{pk}}. \quad (3)$$

Because flare plasma is likely to be, to some degree, multi-thermal (*e.g.* McTiernan *et al.,* 1999; Warren *et al.,* 2013), the isothermally derived peak EM from GOES are treated as a proxy for the EM of the EUV emitting plasma. The key assumption being that the EUV emitting EM is proportional to that of the soft X-ray EM. Eliminating $I_{0,ij}$ from Equation 2 using Equation 3 and collecting terms yields,

$$I_{ij} = \left(G_{ij}(T_{\mathrm{pk}})\mathrm{e}^{-\tau}\right)\mathrm{EM}_{\mathrm{pk}}. \qquad (4)$$

Note, the optical depth of GOES soft X-rays is assumed to be insignificant compared to that of the EUV line emissions because relatively small continua cross-sections dominate the GOES bands. Further, as will be shown later in this section, resonant scattering cross-sections are proportional to the line-center wavelength and inversely proportional to the square-root of the temperature. As such, the resonant scattering cross-sections for any line emissions in the GOES bands are much smaller than those for longer wavelengths.

Equation 4 is rearranged to solve for the *measured* peak contribution function ($G_{\mathrm{M},ij}$) for a set of flares with a mean orthodromic angle, $\theta$, from disk-center,

$$\langle G_{\mathrm{M},ij}(\theta)\rangle \equiv G_{ij}(T_{\mathrm{pk}})\mathrm{e}^{-\langle\tau(\theta)\rangle} = \frac{I_{ij}(T_{\mathrm{pk}})}{\mathrm{EM}_{\mathrm{pk}}}. \qquad (5)$$

In practice, the optical path difference is isolated by finding $G_{\mathrm{M},ij}$ for a set of limb flares with mean angle $\theta_{\mathrm{L}}$ and separately for a set of disk-center flares with mean angle $\theta_{\mathrm{C}}$ via Equation 5 using EVE line irradiance measurements at the time of peak line irradiance and peak GOES EM measurements:

$$\frac{\langle G_{\mathrm{M},ij}(\theta_C)\rangle}{\langle G_{\mathrm{M},ij}(\theta_L)\rangle} = \mathrm{e}^{\Delta\tau}, \qquad (6)$$

where $\Delta\tau = <\tau(\theta_{\mathrm{L}})> - <\tau(\theta_{\mathrm{C}})>$ is the mean optical depth difference.

Next, the optical depth is related to the emission line and plasma parameters. For an emission line, the optical depth for transition $ij$ of species X is related to the transition's resonant scattering cross-section ($\sigma_{ij}$) and ion column density ($N_X$) by

$$\tau_{ij} = \sigma_{ij}N_X. \qquad (7)$$

The resonant scattering cross-section at line-center is related to the line atomic and plasma parameters according to

$$\sigma_{ij} = \frac{g_i}{g_j}A_{ij}\frac{1}{4}\left(\frac{2\pi kT_{X,\mathrm{pk}}}{m_X} + v_{\mathrm{nth}}^2\right)^{-1/2}\lambda_{0,ij}^3 \qquad (8)$$

(*e.g.* Foot 2005), where $m_X$, and $\lambda_{0,ij}$ are the ion mass and line-center wavelength, respectively, $g_i$ and $g_j$ are the degeneracies of the respective levels, $A_{ij}$ is the Einstein A coefficient, $m_X$ is the ion mass, $k$ is Boltzmann's constant and $v_{\mathrm{nth}}$ is the non-thermal ion velocity. Note, other expressions for $\sigma_{ij}$ appear in the literature (e.g. Mariska, 1992; Schrijver and McMullen, 2000) that use the atomic oscillator strength ($f_{ij}$) to show $\sigma_{ij} \propto \lambda_{0,ij}$. This is because $f_{ij} \propto A_{ij}\lambda_{0,ij}^2$. It follows that it is important to treat any dependences of $\sigma_{ij}$ on $\lambda_{0,ij}$ with caution because $A_{ij}$ (or $f_{ij}$) also varies with $\lambda_{0,ij}$. Finally, we collect all terms on the right-hand side of Equation 8 except a factor of $\lambda_{0,ij}$ into a new variable $\xi_{ij}$,

$$\sigma_{ij} = \xi_{ij}\lambda_{0,ij}. \quad (9)$$

The time dependent ion column density is related to the electron column density ($N_e$) by

$$N_X = A_i/A_H F_I N_e = \eta_X N_e, \qquad (10)$$

where $A_i/A_H$ and $F_I$ are the elemental abundance fraction and species ionization fraction, respectively. On the far right-hand side of Equation 10, the expression is simplified by collecting all terms except $N_e$ into a newly defined variable, $\eta_X$.

The factoring of parameters into $\xi_{ij}$ and $\eta_X$ enables a convenient retrieval of the average electron density difference. Equations 7, 9 and 10 are used to re-write Equation 6 as

$$ln\left(\frac{\langle G_{M,ij}(\theta_C)\rangle}{\langle G_{M,ij}(\theta_L)\rangle}\right) = \eta_X \langle \Delta N_e \rangle \xi_{ij} \lambda_{0,ij}, \qquad (11)$$

where $<\Delta N_e> = <N_e (\theta_L, t_{Fe18,pk})> - <N_e (\theta_c, t_{Fe18,pk})>$ is the mean column density difference. Note that Equation 11 inherently assumes constant plasma density over the period which the EUV lines analyzed reach their peak. In reality, this is likely not the case. However, detailed flare emission measure analysis (*e.g.* Ashwanden and Alexander, 2001; Raftery *et al.,* 2009) suggests that the emission measure is relatively constant between 10 MK and 17 MK, and begins to decrease rapidly near the formation temperature of Fe XVIII. As such, the assumption of constant density is reasonable for all lines considered, except possibly Fe XVIII, and will add uncertainty to the analysis. Finally, Equation 11 is rearranged,

$$\xi_{ij}^{-1} \eta_X^{-1} ln\left(\frac{\langle G_{M,ij}(\theta_C)\rangle}{\langle G_{M,ij}(\theta_L)\rangle}\right) = \langle \Delta N_e \rangle \lambda_{0,ij}. \qquad (12)$$

In practice, values of $G_{M,ij}$ are measured for each line at different orthodromic angles and used in Equation 12 to find $<\Delta N_e>$ by the method of least squares.

## 5. Results

Figure 2 shows the peak emission line irradiance and peak soft X-ray EM data used to find $G_{M,ij}$ by solving Equation 4 for three of the five angular subsets considered in this study. The data for the two omitted angular subsets are similar. The columns correspond with different pairs of angular subsets and the rows correspond with the six emission lines considered in this study. The disk-center observations are shown on every panel with blue diamonds, and the red triangles correspond with the limb-ward observations. Blue and red dashed lines show the zero-intercept linear fit for the disk-center and limb-ward data, respectively, where the slope of each line is $G_{M,ij}$ according to Equation 4. Also shown on each panel are the Pearson correlation coefficients, where the text color matches the corresponding symbol color. In general, the peak irradiance and peak EM peak are highly correlated. The correlation increases with increasing line formation temperature as is expected since the soft X-ray peak EM typically corresponds with 15 MK or hotter plasma (and nearest the formation temperature for Fe XXIV) for M-class and larger flares (Feldman and Doschek 1996).

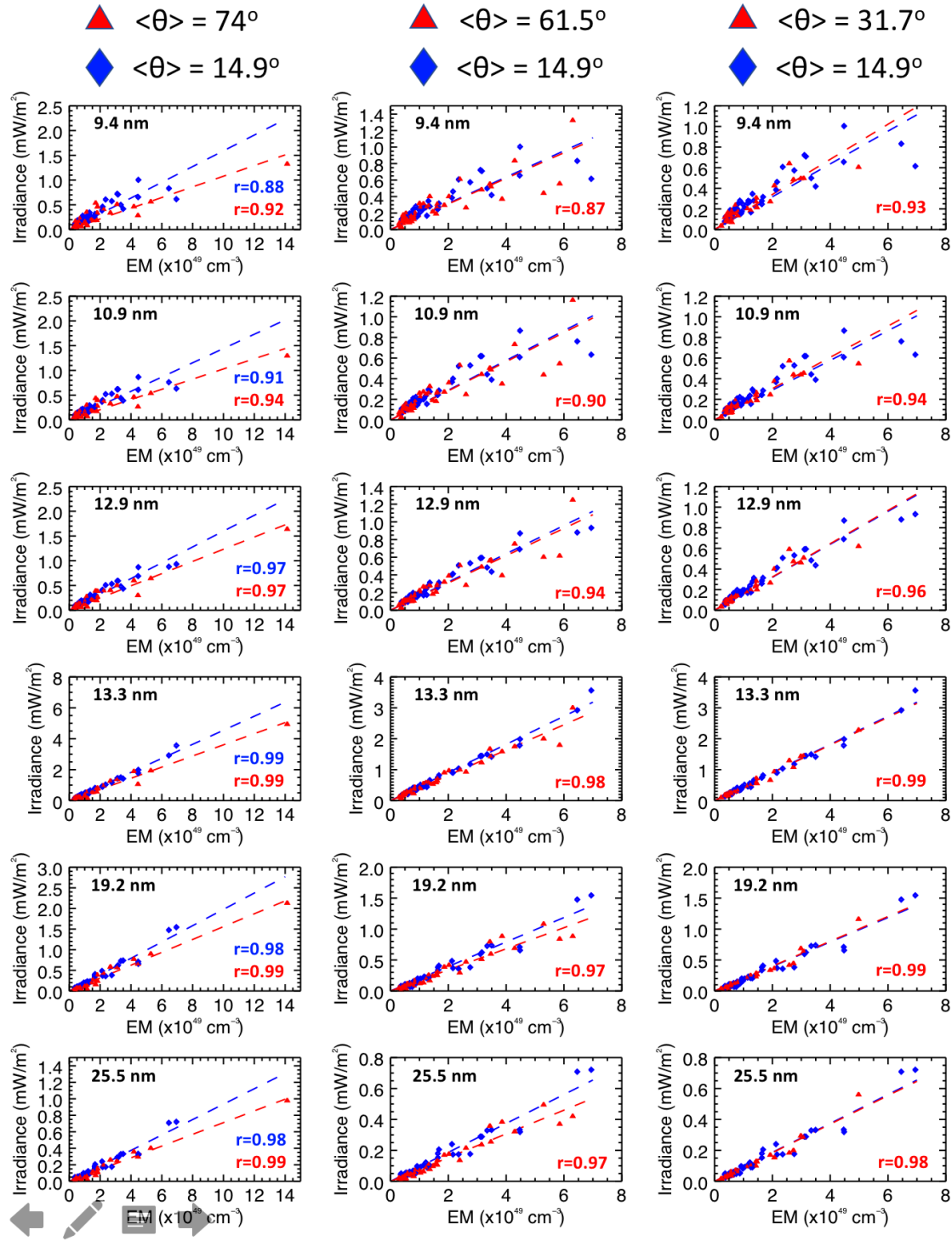

**Figure 2.** EUV irradiance and soft X-ray EM data and first order linear fits used to find $G_{M,ij}(\theta)$ for three of the 5 angle subsets considered in this study. Each column of panels corresponds with a different combination of angular subsets, with the angles and corresponding symbols shown above each column. On each panel, the line-center wavelength and Pearson correlation coefficients between the irradiance and EM are given, where the text color corresponds with the symbol color for the correlation coefficients.

Figure 3 shows results found solving Equation 6 with the $G_{M,ij}(\theta)$ measurements from the five lines and six angles listed in Section 3.2. The subset of 45 flares nearest to disk-center, with a mean angle of 14.9°, are used to find $<G_{M,ij}(\theta_C)>$ and the remaining angular-partitioned subsets are used to find $<G_{M,ij}(\theta_L)>$. The error bars correspond with the standard-deviation of the mean. Panel a shows how $\exp(\Delta\tau)$ varies as a function of angle, with the lines distinguished by differing color. The diamond corresponds with 14.9°, where $\exp(\Delta\tau)$ should equal unity. It is evident that flares that occur near disk-center are systematically brighter than those occurring away from disk-center, with disk-center flare line emissions being, on average, 10-45% brighter than those from flares near the limb. It is important to note that these results show that the observed CTLV is not due to a portion of the loop being over the limb because $\exp(\Delta\tau)$ varies smoothly with decreasing angle from the limb. There is also an approximate trend of linearly decreasing EUV emission intensity with increasing orthodromic angle from disk-center. Panel b shows the same data plotted versus wavelength. It is evident that the two Fe XXIV lines, with wavelengths of 19.2 and 25.5 nm, demonstrate a higher degree of CTLV. There appears to possibly be a weak correlation between the degree of CTLV and wavelength, which is suggested by Equation 8. However, closer inspection shows opposite trends at the shortest and longest wavelengths indicating that the non-wavelength parameters that determine $\sigma_{ij}$ are also important.

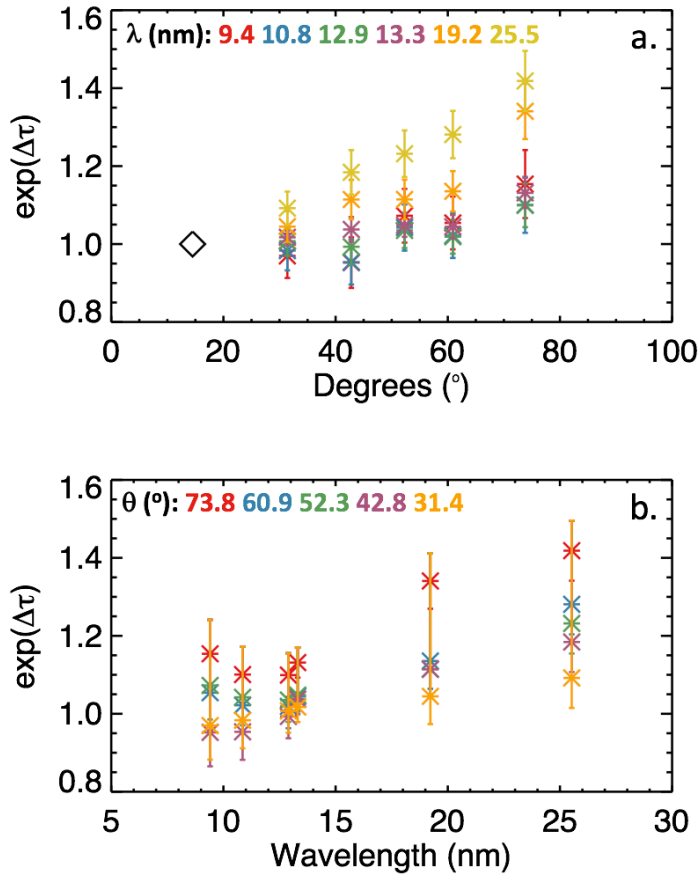

**Figure 3.** Observed brightening of disk-center flares compared to flares located off disk-center. The panels show how the inverse extinction ratio given by Equation 6 varies with angle (a.) and wavelength (b.) for the 5 angles and 6 lines considered.

To investigate the dependence on the atomic physics parameters listed in Table 1, Equation 12 is solved for $<\Delta N_e>$. Results are shown in Figure 4, where each panel corresponds with a different $<\theta_L>$. Fitted values for $<\Delta N_e>$ are shown on each panel, along with $<\theta_L>$ and the number of limb-ward flares ($n_L$) used for each case. For all five cases considered, $\xi_{ij}^{-1}\eta_X^{-1}ln\left(\frac{\langle G_{M,ij}(\theta_C)\rangle}{\langle G_{M,ij}(\theta_L)\rangle}\right)$ varies approximately linearly with wavelength as predicted by Equation 12, although for panels b through d a non-zero intercept would yield a better fit to the data. The resulting values of $<\Delta N_e>$ range from 0.25 x $10^{19}$ to 1.76 x $10^{19}$ cm$^{-3}$, and tend to increase with increasing $<\theta_L>$. The 1-$\sigma$ uncertainties of $<\Delta N_e>$ are also shown and range between 11% and 33% of the best fit value.

The analysis shown in Figure 4 assumes $v_{nth}=0$ although this is likely not the case based on the current understanding and past observations of solar flares, which suggest rapidly flowing plasma at speeds of tens to hundreds of km/s (Tsuneta 1996; Innes, McKenzie and Wang 2003; Mariska 2006). However, using a non-zero value for $v_{nth}$ does not materially change the results and the relative magnitude and any apparent non-linearity in $\xi_{ij}^{-1}\eta_X^{-1}ln\left(\frac{\langle G_{M,ij}(\theta_C)\rangle}{\langle G_{M,ij}(\theta_L)\rangle}\right)$ with wavelength remains unchanged although the resulting values of $<\Delta N_e>$ do increase. For example, assuming $v_{nth}=50$ km/s yields $<\Delta N_e> = 2.67$ x $10^{19}$ cm$^{-2}$ for the $<\theta_L> = 74^o$ case with the fit uncertainty equal to 11%.

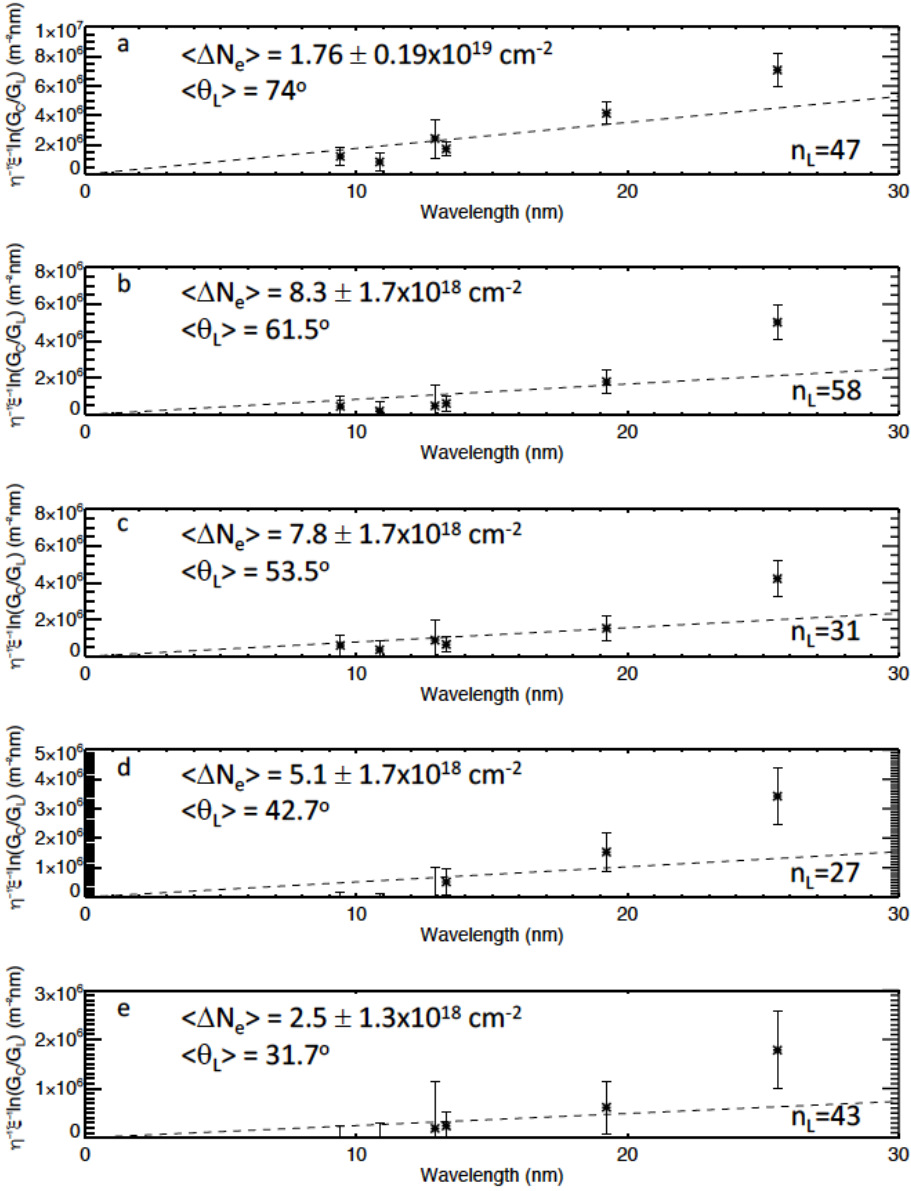

**Figure 4.** Estimates of the average increase in the line-of-site electron column density for off-center versus on-center flares found by solving Equation 12. Each panel corresponds with a different $<\theta_L>$, while $<\theta_C>$ is kept the same (14.5°) for each case. The linear regression coefficient corresponds with $<\Delta N_e>$ and is shown on each plot along with the $<\theta_L>$ and number of flares ($n_L$) in the subsets considered.

Figure 5 plots the best-fit values for $<\Delta N_e>$ versus $<\theta_L>$ using asterisks. $<\Delta N_e>$ is constrained to be equal to zero when $<\theta_L>$ is equal to $<\theta_C>$ as is indicated with the diamond. The dashed curve shown in Figure 5 corresponds with the best-fit equation,

$$< \Delta N_e >= 0.022\theta_L - 0.32, \quad (12)$$

to the data. Although Equation (12) is empirically determined and provides little physical insight, it is useful for estimating $<\Delta N_e>$, which is important for understanding and predicting the CTLV of a flare spectrum as will be shown in Section 6.

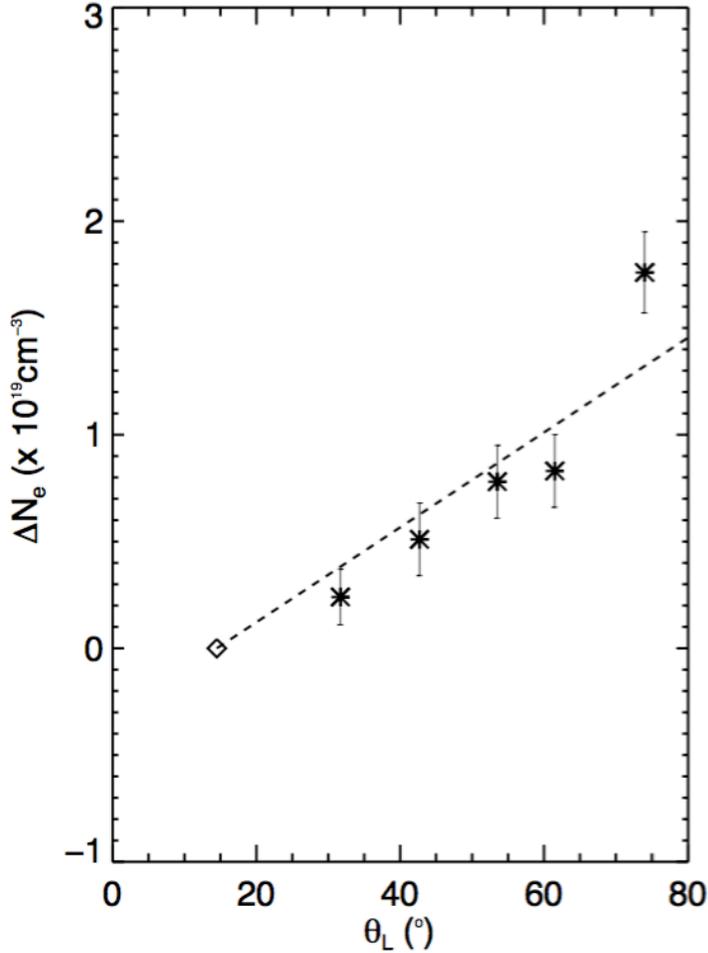

**Figure 5**. Variation of $<\Delta N_e>$ with $<\theta_L>$. The data are fit to a logarithmic function given by Equation (12).

## 6. Discussion

The spatial scale of the obscuring volume of plasma, $\Delta z$, is estimated from the $<\Delta N_e(73.8°)> = 1.8 \times 10^{19}$ cm$^{-3}$ value inferred in Section 5 using the approximation $<\Delta N_e> = n_e \Delta z$: Assuming $n_e$ in the range of $10^{11}$-$10^{12}$ (Milligan *et al.*, 2012), $\Delta z$ is found to be in the range of 0.18 - 1.8 Mm. This range of values is at the bottom end of as estimates of flare loop diameters, which have been found to range from 1.5 to 6 Mm (Aschwanden *et al.*, 2000). This is to be expected when considering that the preceding simple spatial scale estimate assumes that the entire volume of plasma is obscured equally, when in reality it is not. For example, the configuration in Figure 1c results in near-total extinction of the loop leg farthest from the observer, but little extinction of the loop leg nearest the observer. As such, the derived spatial scales will be biased small and are likely only useful for order-of-magnitude estimates. Further, the assumption of $v_{nth}=0$ implies that $1.8 \times 10^{19}$ cm$^{-3}$ is a lower bound of $<\Delta N_e(73.8°)>$, also biasing the estimates of $\Delta z$ low. Additionally, Figure 1c almost certainly oversimplifies the scenario. In reality, a flare loop is not likely oriented perfectly equatorially (even though Joy's law implies there will likely be some equatorial component), and any deviation from an equatorial orientation will reduce the amount of a loop that is self-obscured. Future studies using three dimensional magnetohydrodynamic models of solar flare loops should enable further progress towards understanding the factors contributing to the observed CTLV.

The optically thick nature of solar flare EUV line emissions has consequences for both estimating flare irradiance and spectroscopically deriving physical properties of flare plasma, such as temperature and density, from available measurements. These consequences are investigated by simulating how the EUV spectrum evolves for a limb flare observed by GOES XRS using the CHIANTI database and the $<\Delta N_e(73.8°)>$ value inferred in Section 5, where CHIANTI is used to both synthesize flare spectra and compute $\Delta\tau_{ij}$.

The flare used to drive the simulation is an M6.9 flare, occurring on 8 July 2012. Inputs and results of this simulation are shown in Figure 6. Panels a and b show the GOES-Long flux and flare temperature for the flare considered. The flare temperature, initially found by the instrument response and spectral contribution functions described by White *et al.* (2005), is decreased by a factor of 0.71 to account for the average hot bias between GOES and AIA found by Ryan *et al.* (2014). This temperature and the GOES XRS derived emission measure are used as inputs to CHIANTI to compute the corresponding isothermal solar flare spectrum. The emission measure is also used to scale $<\Delta N_e>$, which is taken to be equal to be 1.8 x $10^{19}$ cm$^{-3}$ at the EM peak and vary proportionally to $\sqrt{\text{EM}}$ otherwise. For each line, $\Delta N_e(t)$ is used to find $\Delta\tau_{ij}$, which in-turn is used to find how much the line is attenuated relative to the disk-center spectrum using the Beer-Lambert law. Note that only the intensity of lines with peak formation temperatures at or above 1.3 MK are modified to account for CTLV because lines that form below this temperature can exhibit coronal dimming (Woods *et al.*, 2011), which may have different CTLV. CTLV of cooler forming chromospheric and transition region lines are also expected to cause limb darkening. As such, this simulation captures only the hot coronal component of spectral CTLV effects.

Figure 6c shows the simulated decrease in limb flare irradiance relative to irradiance from the same flare observed near disk-center for two bands of aeronomic interest; the solid curve corresponds with the 10-40 nm wavelength range, while the dashed curve corresponds with the 0.1-100 nm range. It is evident that due to CTLV in hot coronal lines alone, the limb flare is substantially dimmer in both bands over the course of the flare, with the limb flare being near 10% (30%) for the 0.1-100 nm (10-40 nm) range. The irradiance difference rapidly increases as the loop density increases, reaching a maximum at the time of peak EM. The difference then begins to decrease with decreasing flare density, but reverses course and increases again after approximately 1300 s. This later increase corresponds with when cooler forming, longer wavelength emissions that have larger resonant scattering cross-sections reach their temperatures of peak formation. This is confirmed by inspecting how the simulated $\Delta\tau_{ij}$ values evolve during the flare, a sample of which is shown in Figure 8, where the three panels correspond with three different times during the flare. The corresponding time is printed in each panel. Notice that $\Delta\tau_{ij}$ for emissions that peak later in flare (Panel c) are 5-10 times greater than those occurring during the rise and peak of the flare (Panels a and b).

Measurements from multiple AIA bands are frequently combined to estimate the plasma temperature, emission measure and differential emission measure (*e.g.* Schmelz *et al.,* 2011; Aschwanden *et al.,* 2013). Therefore, it is important to understand the relative CTLV that is expected in the AIA bands. Figure 6d shows the simulated limb darkening for the five temperature bands with peak temperature sensitivity above 1.3 MK. The 21.1 nm and 19.3 nm bands have comparable CTLV after approximately 600 s, when they are sensitive to the relatively cool plasma emitting during the later flare phase. However, the 19.3 nm band shows over 20% limb darkening preceding the flare soft X-ray peak, when the Fe XXIV emission line dominates the pass band. The limb darkening for the 13.1 nm and 33.5 nm bands evolves comparably until after approximately 1300 s, when the limb darkening in the 33.5 nm band begins to increase and eventually exceed 40%. And the limb darkening in the 9.4 nm band differs from all of the other bands, gradually increasing throughout the flare and reaching 40% during the flare declining phase. It follows that quantities (density, temperature, DEM, etc.) derived from the AIA bands for this simulated flare will be different if the flare occurs at the limb versus disk-center, and care should be taken when using AIA images to characterize flare plasma to ensure the optically thin approximation is valid for the regions of the images being considered. It should be noted that the brightest loops of large flares typically saturate the AIA detector, preventing them from being combined for flare spectral analyses regardless of CTLV effects.

Measurements of EVE line ratios have been used to estimate flare density using the optically thin assumption (Milligan *et al.,* 2012), and different CTLV in the lines used will result in different derived temperatures, depending on whether the flare is observed at disk-center or the limb. The Fe XXI 14.573 nm to 12.875 nm line ratio is sensitive to plasma density in the $10^{11}$-$10^{13}$ cm$^{-3}$ range, but differences in the resonant scattering cross-sections of the line causes CTLV in the line ratio. Figure 6e shows simulation results for how the 14.573/12.875 ratio differs when the flare is observed at the limb from when it is observed at disk-center. The maximum differences are approximately +/- 10%. For densities near $3 \times 10^{11}$ cm$^{-3}$, a 10% error in the line ratio results in a 15-20% error in the derived densities.

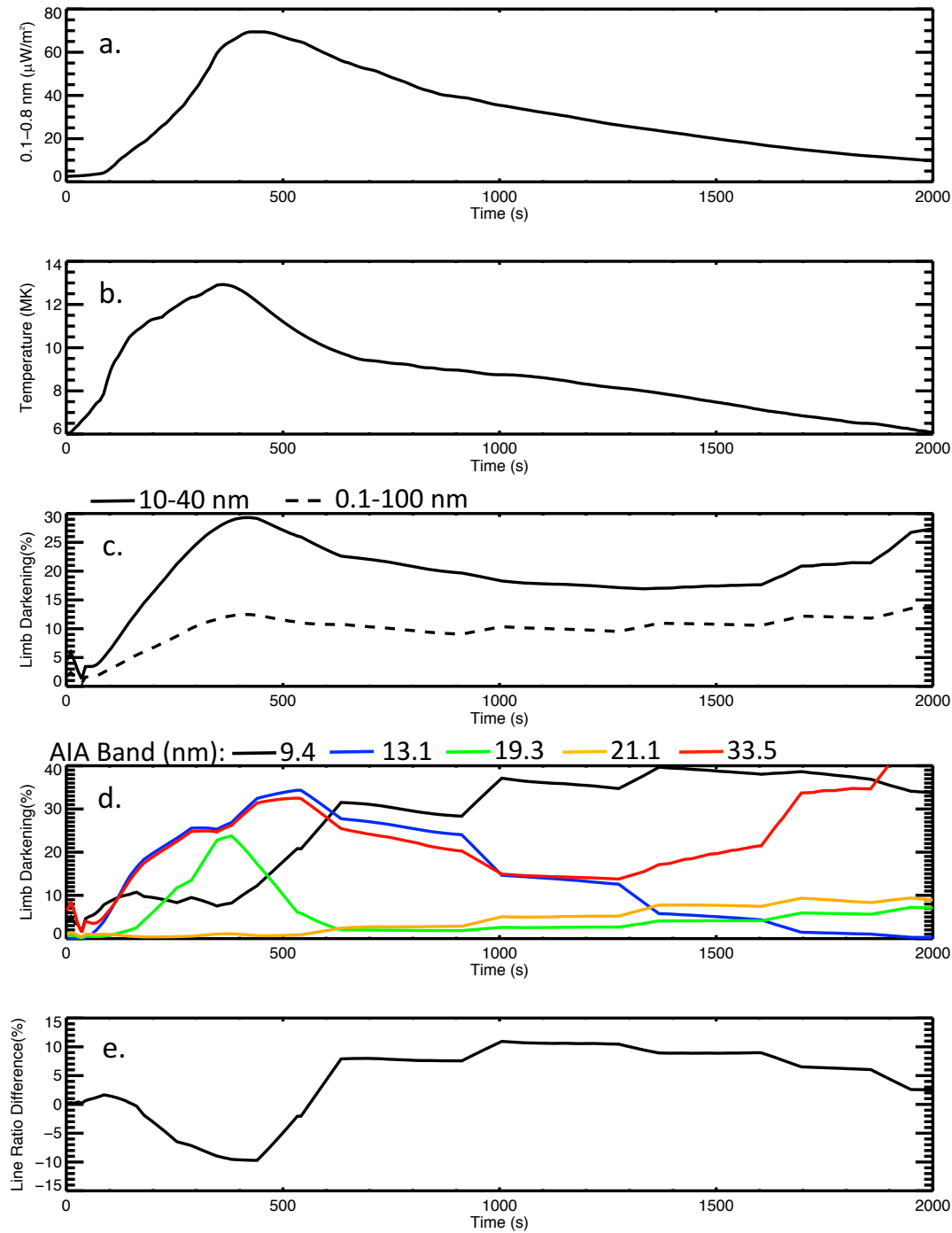

**Figure 6.** Simulated effects of flare CTLV on sample measurements of aeronomic and solar physics interest. a) The 0.1-0.8 nm measured GOES XRS irradiance and b) the derived isothermal temperature used to drive the simulation. c) The percent decrease in integrated 10-40 nm and 0.1-100 nm irradiance for the flare occurring at the limb compared to near disk-center. d) Similar to c but for the five hottest AIA bands. e) Difference in the Fe XXI 14.573nm/12.875nm line ratio for the flare occurring at the limb from its value when observed near disk-center.

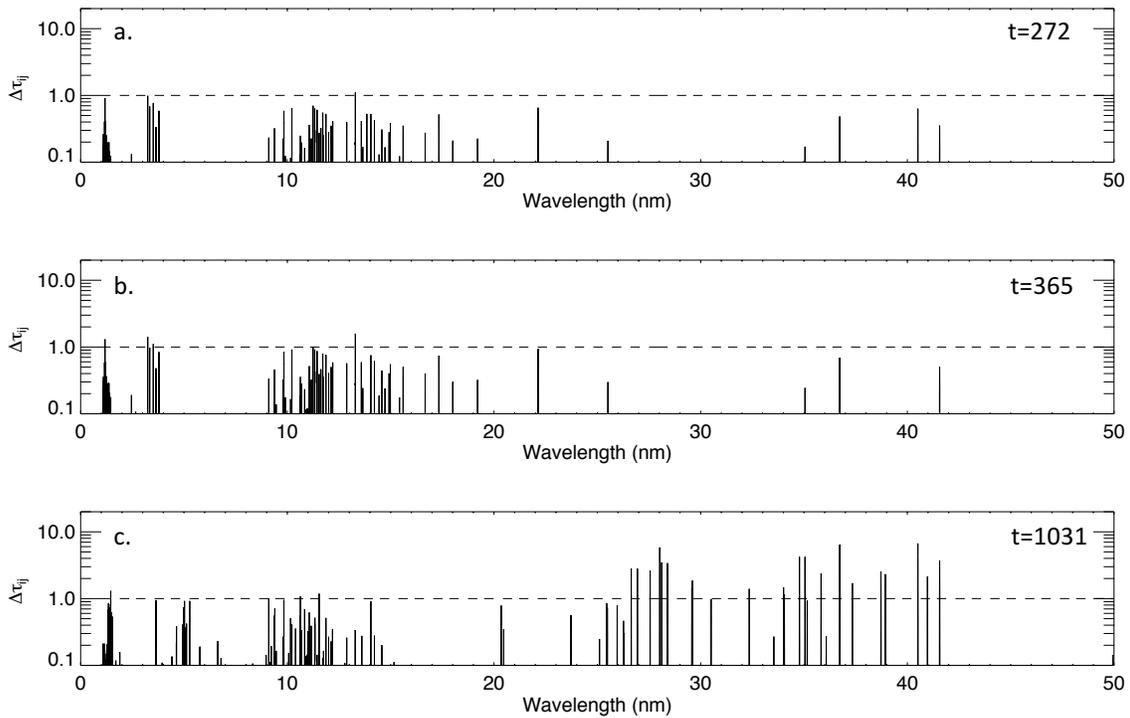

**Figure 7.** Increases in optical density for the simulated flare in Figure 6 at five different times. The time indicated on each panel corresponds with the seconds shown on the x axes of Figure 6.

## 6. Conclusions

i)      CTLV in hot coronal emissions is expected based on the current understanding of flaring loop densities and geometries. Specifically, emissions with formation temperatures near 10 MK will limb darken. For example, observations show that observed flare irradiance will be approximately 10% smaller for flares located 45° from disk-center.

ii)     Flare CTLV can be used to estimate the increase in column density along the line-of-sight for limb flares versus disk-center flares. This analysis of EVE flare measurements shows the typical increase in column density for limb flares to be $1.76 \times 10^{19}$ cm$^{-2}$. This column density increase is consistent with that of a loop diameter, suggesting that limb darkening occurs as further (with respect to the observer) portions of a loop are obscured by nearer portions of the loop, an effect that is exacerbated as the loop approaches the limb.

iii)    The optically thin approximation is not valid for analyses of M-class or larger flares; further study is needed to see if this is true for smaller flares as well. The optically thin approximation becomes increasingly invalid for flares located further away from disk-center, where portions of the flaring loop can be obscured by high density plasma. Because resonant scattering cross-sections vary substantially from line-to-line, the degree of limb darkening is not uniform as a function of wavelength. As such, analyses that rely on relative measurements, such as line ratios, will be subject to error if the plasma is assumed optically thin.

iv)      CTLV of hot coronal emissions can cause a substantial reduction in ionizing radiation impacting planetary atmospheres.  Further study is needed to compare the relative magnitudes of CTLV occurring in the transition region and chromosphere with CTLV occurring in the corona.

## 7. Acknowledgements

This study was funded by NASA Living with a Star Grant NNX16AE86G titled, "Improving Solar EUV Spectral Irradiance Models with Multi-Vantage Point Observations".

This study used the CHIANT database; CHIANTI is a collaborative project involving George Mason University, the University of Michigan (USA) and the University of Cambridge (UK).

The authors would like to thank Drs A. Kowalski of the National Solar Observatory, T.N. Woods, A.R. Jones, D.W. Woodraska of the Laboratory for Atmospheric and Space Physics, and M. West of the Royal Observatory of Belgium for the helpful conversations regarding various aspects of this study.

## 8. Declaration of Potential Conflicts of Interest

None.

# 9. Appendix

Table A1.  List of flare dates, times, GOES XRS magnitudes and center-to-limb angles.

| YearDay | Hour | Mag. | Angle (°) | YearDay | Hour | Mag. | Angle (°) | YearDay | Hour | Mag. | Angle (°) | YearDay | Hour | Mag. | Angle (°) | YearDay | Hour | Mag. | Angle (°) |
|---|---|---|---|---|---|---|---|---|---|---|---|---|---|---|---|---|---|---|---|
| 2011047 | 14.4 | M1.6 | 31 | 2012019 | 16 | M3.2 | 42 | 2012209 | 17.5 | M2.7 | 63 | 2013298 | 10.2 | M1.0 | 72 | 2014062 | 16 | M1.2 | 37 |
| 2011066 | 7.9 | M1.4 | 63 | 2012023 | 4 | M8.7 | 39 | 2012210 | 21 | M6.1 | 52 | 2013298 | 15.1 | X2.1 | 67 | 2014069 | 4.1 | M1.0 | 47 |
| 2011066 | 9.3 | M1.8 | 50 | 2012027 | 18.6 | X1.7 | 63 | 2012211 | 6.4 | M2.3 | 48 | 2013298 | 21 | M1.9 | 63 | 2014069 | 15.5 | M1.7 | 40 |
| 2011066 | 14.5 | M1.9 | 26 | 2012037 | 19.9 | M1.0 | 60 | 2012212 | 15.8 | M1.1 | 36 | 2013299 | 6.1 | M2.3 | 59 | 2014070 | 12.1 | M1.7 | 65 |
| 2011066 | 20.2 | M3.7 | 53 | 2012062 | 17.8 | M3.3 | 73 | 2012219 | 4.6 | M1.6 | 76 | 2013299 | 19.5 | M3.1 | 75 | 2014071 | 22.6 | M9.3 | 68 |
| 2011067 | 2.3 | M1.3 | 69 | 2012064 | 10.9 | M2.0 | 60 | 2012224 | 12.3 | M1.0 | 46 | 2013301 | 2 | X1.0 | 71 | 2014079 | 19.3 | M1.2 | 74 |
| 2011067 | 3.9 | M1.5 | 63 | 2012065 | 4.2 | X1.1 | 57 | 2012230 | 13.3 | M2.4 | 73 | 2013301 | 4.7 | M5.1 | 69 | 2014079 | 3.9 | M1.7 | 35 |
| 2011067 | 10.8 | M5.3 | 71 | 2012065 | 19.3 | M2.1 | 46 | 2012231 | 1 | M5.5 | 71 | 2013302 | 21.9 | X2.3 | 4 | 2014087 | 19.3 | M2.0 | 26 |
| 2011067 | 18.5 | M4.4 | 71 | 2012065 | 19.5 | M1.8 | 46 | 2012231 | 3.4 | M1.8 | 71 | 2013304 | 13.8 | M1.9 | 9 | 2014088 | 17.8 | X1.0 | 35 |
| 2011067 | 20.3 | M1.4 | 71 | 2012069 | 3.9 | M6.3 | 23 | 2012231 | 22.9 | M1.0 | 69 | 2013306 | 22.4 | M1.6 | 19 | 2014090 | 8.1 | M1.4 | 70 |
| 2011071 | 4.7 | M1.3 | 37 | 2012070 | 17.8 | M8.4 | 30 | 2012250 | 4.2 | M1.6 | 41 | 2013307 | 5.4 | M5.0 | 22 | 2014092 | 14.1 | M6.5 | 51 |
| 2011073 | 19.9 | M4.2 | 49 | 2012074 | 15.4 | M2.8 | 21 | 2012252 | 18 | M1.4 | 41 | 2013311 | 14.5 | M2.4 | 27 | 2014108 | 13 | M7.3 | 34 |
| 2011074 | 0.4 | M1.0 | 71 | 2012077 | 20.7 | M1.3 | 27 | 2012274 | 4.6 | M1.3 | 75 | 2013312 | 4.4 | X1.1 | 24 | 2014115 | 0.5 | X1.3 | 74 |
| 2011082 | 2.3 | M1.4 | 66 | 2012107 | 17.8 | M1.7 | 74 | 2012294 | 18.3 | M9.0 | 82 | 2013314 | 5.3 | X1.1 | 22 | 2014126 | 9 | M1.8 | 74 |
| 2011083 | 12.1 | M1.0 | 43 | 2012118 | 8.4 | M1.0 | 32 | 2012295 | 20 | M1.3 | 71 | 2013315 | 11.3 | M2.4 | 66 | 2014127 | 16.5 | M1.2 | 80 |
| 2011084 | 23.4 | M1.0 | 25 | 2012126 | 13.4 | M1.4 | 75 | 2012296 | 18.9 | M5.0 | 61 | 2013320 | 4.9 | M1.2 | 34 | 2014128 | 10.1 | M5.2 | 55 |
| 2011105 | 17.2 | M1.3 | 28 | 2012126 | 23 | M1.3 | 71 | 2012297 | 3.3 | X1.8 | 57 | 2013323 | 10.5 | X1.0 | 65 | | | | |
| 2011112 | 15.9 | M1.2 | 36 | 2012128 | 14.5 | M1.9 | 40 | 2012313 | 2.4 | M1.7 | 76 | 2013325 | 11.2 | M1.0 | 76 | | | | |
| 2011148 | 21.9 | M1.1 | 65 | 2012129 | 13.1 | M1.4 | 45 | 2012316 | 2.5 | M1.0 | 70 | 2013327 | 2.6 | M1.1 | 56 | | | | |
| 2011149 | 10.5 | M1.4 | 60 | 2012130 | 12.5 | M4.7 | 34 | 2012318 | 2.1 | M6.0 | 47 | 2013341 | 7.5 | M1.2 | 48 | | | | |
| 2011158 | 6.7 | M2.5 | 51 | 2012130 | 14.2 | M1.8 | 24 | 2012318 | 20.9 | M2.8 | 37 | 2013354 | 12 | M1.6 | 71 | | | | |
| 2011165 | 21.8 | M1.3 | 71 | 2012130 | 21.1 | M4.1 | 33 | 2012325 | 19.5 | M1.6 | 16 | 2013363 | 7.9 | M3.1 | 14 | | | | |
| 2011211 | 2.2 | M9.3 | 36 | 2012131 | 4.3 | M5.7 | 26 | 2012326 | 6.9 | M1.4 | 11 | 2014001 | 18.9 | M9.9 | 46 | | | | |
| 2011214 | 6.3 | M1.4 | 15 | 2012131 | 20.4 | M1.7 | 20 | 2012332 | 16 | M1.6 | 71 | 2014002 | 2.6 | M1.7 | 77 | | | | |
| 2011215 | 3.6 | M1.1 | 27 | 2012138 | 1.8 | M5.1 | 76 | 2013013 | 0.8 | M1.0 | 27 | 2014003 | 12.8 | M1.0 | 50 | | | | |
| 2011215 | 13.8 | M6.0 | 29 | 2012155 | 17.9 | M3.3 | 40 | 2013013 | 8.6 | M1.7 | 29 | 2014004 | 10.4 | M1.3 | 47 | | | | |
| 2011220 | 18.2 | M3.5 | 58 | 2012158 | 20.1 | M2.1 | 18 | 2013095 | 17.8 | M2.2 | 6 | 2014007 | 3.9 | M1.0 | 13 | | | | |
| 2011221 | 3.9 | M2.5 | 62 | 2012161 | 16.9 | M1.8 | 68 | 2013101 | 7.2 | M6.5 | 18 | 2014007 | 10.2 | M7.2 | 16 | | | | |
| 2011248 | 8 | M1.2 | 73 | 2012162 | 6.8 | M1.3 | 63 | 2013102 | 20.6 | M3.3 | 44 | 2014007 | 18.5 | X1.2 | 12 | | | | |
| 2011249 | 1.8 | M5.3 | 9 | 2012165 | 13.2 | M1.2 | 26 | 2013122 | 5.2 | M1.1 | 28 | 2014008 | 3.8 | M3.6 | 76 | | | | |
| 2011249 | 22.4 | X2.1 | 18 | 2012166 | 14.4 | M1.9 | 20 | 2013123 | 16.9 | M1.3 | 41 | 2014013 | 21.9 | M1.3 | 73 | | | | |
| 2011250 | 22.7 | X1.8 | 30 | 2012180 | 16.2 | M2.4 | 45 | 2013123 | 17.5 | M5.7 | 73 | 2014027 | 2.1 | M1.0 | 74 | | | | |
| 2011251 | 15.8 | M6.7 | 39 | 2012182 | 12.9 | M1.0 | 23 | 2013130 | 1 | M3.9 | 14 | 2014028 | 19.6 | M4.9 | 69 | | | | |
| 2011253 | 7.7 | M1.1 | 61 | 2012182 | 18.5 | M1.6 | 21 | 2013133 | 2.3 | X1.7 | 85 | 2014028 | 22.3 | M2.6 | 67 | | | | |
| 2011266 | 23.9 | M1.9 | 62 | 2012183 | 19.3 | M2.8 | 12 | 2013135 | 1.8 | X1.2 | 61 | 2014030 | 8.2 | M1.1 | 51 | | | | |
| 2011267 | 17.4 | M1.7 | 67 | 2012184 | 0.6 | M1.1 | 12 | 2013136 | 21.9 | M1.3 | 43 | 2014031 | 15.7 | M1.1 | 35 | | | | |
| 2011268 | 17 | M2.2 | 42 | 2012184 | 10.9 | M5.6 | 21 | 2013142 | 13.5 | M5.0 | 71 | 2014032 | 6.6 | M2.6 | 25 | | | | |
| 2011273 | 19.1 | M1.0 | 5 | 2012184 | 20.1 | M3.8 | 19 | 2013151 | 20 | M1.0 | 43 | 2014033 | 8.4 | M2.2 | 14 | | | | |
| 2011294 | 13 | M1.3 | 75 | 2012186 | 16.7 | M1.8 | 33 | 2013156 | 9 | M1.3 | 55 | 2014033 | 18.2 | M3.1 | 9 | | | | |
| 2011306 | 22 | M4.3 | 67 | 2012186 | 22.2 | M4.6 | 32 | 2013158 | 22.9 | M5.9 | 51 | 2014035 | 4 | M5.2 | 10 | | | | |
| 2011307 | 11.2 | M2.5 | 59 | 2012186 | 23.9 | M1.2 | 32 | 2013172 | 3.2 | M2.9 | 70 | 2014038 | 5 | M2.0 | 49 | | | | |
| 2011307 | 20.5 | X1.9 | 57 | 2012187 | 1.2 | M2.4 | 32 | 2013224 | 10.7 | M1.5 | 31 | 2014042 | 16.9 | M1.8 | 9 | | | | |
| 2011309 | 3.6 | M3.7 | 44 | 2012187 | 2.7 | M2.2 | 33 | 2013229 | 18.4 | M3.3 | 32 | 2014043 | 4.4 | M3.7 | 6 | | | | |
| 2011310 | 1.2 | M1.2 | 34 | 2012187 | 7 | M1.1 | 35 | 2013229 | 19.5 | M1.4 | 32 | 2014045 | 3 | M2.3 | 24 | | | | |
| 2011313 | 13.6 | M1.1 | 35 | 2012187 | 13.3 | M1.2 | 43 | 2013282 | 1.8 | M2.8 | 43 | 2014045 | 12.7 | M1.6 | 34 | | | | |
| 2011319 | 9.2 | M1.2 | 64 | 2012188 | 1.7 | M2.9 | 42 | 2013286 | 0.7 | M1.7 | 32 | 2014047 | 9.4 | M1.1 | 3 | | | | |
| 2011319 | 22.6 | M1.1 | 69 | 2012188 | 13.5 | M1.2 | 72 | 2013296 | 20.9 | M2.7 | 1 | 2014051 | 8 | M3.0 | 68 | | | | |
| 2011359 | 18.3 | M4.0 | 31 | 2012188 | 23.1 | X1.1 | 59 | 2013296 | 24.5 | M1.4 | 9 | 2014055 | 12.1 | M1.2 | 76 | | | | |
| 2011360 | 2.5 | M1.5 | 35 | 2012189 | 3.3 | M1.2 | 54 | 2013296 | 24.5 | M3.1 | 8 | 2014056 | 1 | X4.9 | 71 | | | | |
| 2011363 | 13.8 | M1.9 | 59 | 2012190 | 16.6 | M6.9 | 72 | 2013297 | 10.5 | M2.5 | 11 | 2014057 | 15 | M1.1 | 43 | | | | |
| 2011363 | 21.9 | M2.0 | 58 | 2012191 | 23.1 | M1.1 | 40 | 2013297 | 10.6 | M3.5 | 11 | 2014059 | 0.8 | M1.1 | 49 | | | | |
| 2011364 | 3.2 | M1.2 | 57 | 2012194 | 16.8 | X1.4 | 18 | 2013298 | 3.1 | M2.9 | 74 | 2014060 | 13.6 | M1.1 | 75 | | | | |
| 2012017 | 4.9 | M1.0 | 52 | 2012201 | 6 | M7.7 | 15 | 2013298 | 8 | X1.7 | 71 | 2014061 | 23.3 | M1.1 | 68 | | | | |

# 10. References

Aschwanden, M. J., Alexander, D.: 2001, Flare plasma cooling from 30 MK down to 1 MK modeled from Yohkoh, GOES, and TRACE observations during the Bastille day event (14 July 2000). *Solar Physics*. **204**, 1. DOI: 10.1023/A:1014257826116.

Aschwanden, M. J., Nightingale, R. W., Alexander, D.: 2000, Evidence for nonuniform heating of coronal loops inferred from multithread modeling of TRACE data. *The Astrophysical Journal*. **541**, 2. DOI: 10.1086/309486.

Aschwanden, M. J., Boerner, P., Schrijver, C. J., Malanushenko, A.: 2013, Automated temperature and emission measure analysis of coronal loops and active regions observed with the Atmospheric Imaging Assembly on the Solar Dynamics Observatory (SDO/AIA). *Solar Physics*. **283**, 1. DOI: 10.1007/s11207-011-9876-5.

Bornmann, P. L., Speich, D., Hirman, J., Matheson, L., Grubb, R., Garcia, H. A., Viereck, R.: 1996, October, GOES X-ray sensor and its use in predicting solar-terrestrial disturbances. In *SPIE's 1996 International Symposium on Optical Science, Engineering, and Instrumentation* (pp. 291-298). International Society for Optics and Photonics. DOI: 10.1117/12.254076.

Chamberlin, P. C., Woods, T. N., Eparvier, F. G.: 2008, Flare irradiance spectral model (FISM): Flare component algorithms and results. *Space Weather*, **6**, 5. DOI: 10.1029/2007SW000372.

Del Zanna, G., Dere, K. P., Young, P. R., Landi, E., Mason, H. E.: 2015, CHIANTI–An atomic database for emission lines. Version 8. *Astronomy & Astrophysics*. **582**, A56. DOI: 10.1051/aas:1997368.

Feldman, U., Doschek, G. A., Behring, W. E., Phillips, K. J. H.: 1996, Electron temperature, emission measure, and X-ray flux in A2 to X2 X-ray class solar flares. *The Astrophysical Journal*. **460**, 1034. DOI: 10.1086/177030.

Freeland, S. L., Handy, B. N.: 1998, Data analysis with the SolarSoft system. *Solar Physics*. **182**, 2. DOI: 10.1023/A:100503822.

Foot, C. J.: 2005. *Atomic physics* (Vol. 7). Oxford University Press.

Hock, R. A.: 2012, *The role of solar flares in the variability of the extreme ultraviolet solar spectral irradiance* (Doctoral dissertation, University of Colorado at Boulder).

Innes, D. E., McKenzie, D. E., Wang, T.: 2003, Observations of 1000 km s− 1 Doppler shifts in 10 7 K solar flare supra-arcade. *Solar Physics*. **217**, 2. DOI: 10.1023/B:SOLA.0000006874.31799.bc.

Lemen, J. R., Akin, D. J., Boerner, P. F., Chou, C., Drake, J. F., Duncan, D. W., ... Katz, N. L.: 2011, The atmospheric imaging assembly (AIA) on the solar dynamics observatory (SDO). In *The Solar Dynamics Observatory* (pp. 17-40). Springer US. DOI: https://doi.org/10.1007/978-1-4614-3673-7_3.

Mariska, J. T.: 1992, *The solar transition region* (Vol. 23). Cambridge University Press. DOI.

Mariska, J. T.: 2006, Characteristics of solar flare Doppler-shift oscillations observed with the Bragg Crystal Spectrometer on Yohkoh. *The Astrophysical Journal*. **639**, 1. DOI: 10.1086/499296

Mason, H. E., Doschek, G. A., Feldman, U., Bhatia, A. K.: 1979, Fe XXI as an electron density diagnostic in solar flares. *Astronomy and Astrophysics*. **73**, 74-81.

McTiernan, J. M., Petrosian, V.: 1991, Center-to-limb variations of characteristics of solar flare hard X-ray and gamma-ray emission. *The Astrophysical Journal*. **379**, 381-391. DOI: 10.1086/170513.

McTiernan, J. M., Fisher, G. H., Li, P.: 1999, The solar flare soft X-ray differential emission measure and the Neupert effect at different temperatures. *The Astrophysical Journal*. **514**, 1. DOI: 10.1086/306924.

Milligan, R. O., Kennedy, M. B., Mathioudakis, M., Keenan, F. P.: 2012, Time-dependent density diagnostics of solar flare plasmas using SDO/EVE. *The Astrophysical Journal Letters*. **755**, 1. DOI: 10.1086/306924.

Raftery, C. L., Gallagher, P. T., Milligan, R. O., Klimchuk, J. A.: 2009, Multi-wavelength observations and modelling of a canonical solar flare. *Astronomy & Astrophysics*. **494**, 3. DOI: 10.1051/0004-6361:200810437.

Ryan, D. F., O'Flannagain, A. M., Aschwanden, M. J., Gallagher, P. T.: 2014, The compatibility of flare temperatures observed with AIA, GOES, and RHESSI. *Solar Physics*. **289**, 7. DOI: https://doi.org/10.1007/s11207-014-0492-z.

Schmelz, J. T., Jenkins, B. S., Worley, B. T., Anderson, D. J., Pathak, S., Kimble, J. A.: 2011, Isothermal and multithermal analysis of coronal loops observed with AIA. *The Astrophysical Journal*. **731**, 1. DOI: 10.1088/0004-637X/731/1/49.

Schrijver, C. J., McMullen, R. A.: 2000, A case for resonant scattering in the quiet solar corona in extreme-ultraviolet lines with high oscillator strengths. *The Astrophysical Journal*. **531**, 2. DOI: 10.1086/308497.

Thomas, R. J., Starr, R., Crannell, C. J.: 1985, Expressions to determine temperatures and emission measures for solar X-ray events from GOES measurements. *Solar physics*. **95**, 2. DOI: 10.1007/BF00152409.

Tsuneta, S.: 1996, Structure and dynamics of magnetic reconnection in a solar flare. *The Astrophysical Journal*. **456**, 840. DOI: 10.1086/176701.

Warren, H. P., Mariska, J. T., Doschek, G. A.: 2013, Observations of thermal flare plasma with the EUV variability experiment. *The Astrophysical Journal*. **770**, 2. DOI: 10.1088/0004-637X/770/2/116.

White, S. M., Thomas, R. J., Schwartz, R. A.: 2005, Updated expressions for determining temperatures and emission measures from GOES soft X-ray measurements. *Solar Physics*. **227**, 2. DOI: 10.1007/s11207-005-2445-z.

Woods, T. N., Eparvier, F. G., Bailey, S. M., Chamberlin, P. C., Lean, J., Rottman, G. J., ... Woodraska, D. L.: 2005, Solar EUV Experiment (SEE): Mission overview and first results. *Journal of Geophysical Research: Space Physics*. **110**, A1. DOI: 10.1029/2004JA010765.

Woods, T. N., Eparvier, F. G., Hock, R., Jones, A. R., Woodraska, D., Judge, D., ... McMullin, D.: 2010, Extreme Ultraviolet Variability Experiment (EVE) on the Solar Dynamics Observatory (SDO): Overview of science objectives, instrument design, data products, and model developments. In *The Solar Dynamics Observatory* (pp. 115-143). Springer US. DOI: 10.1007/978-1-4614-3673-7_7.

Woods, T. N., Hock, R., Eparvier, F., Jones, A. R., Chamberlin, P. C., Klimchuk, J. A., ... Schrijver, C. J.: 2011, New solar extreme-ultraviolet irradiance observations during flares. *The Astrophysical Journal*. **739**, 2. DOI: 10.1088/0004-637X/739/2/59.